# Solenoid from experimental HTS tape for magnetic refrigeration.

E.P. Krasnoperov, V.V. Guryev, S.V. Shavkin, V.E. Krylov, V.V. Sychugov,

V.S. Korotkov, A.V. Ovcharov, P.V. Volkov

National Research Center "Kurchatov Institute", Kurchatov sq., 1, Moscow, 123182, Russia

*Abstract.*

*The 3-Tesla superconducting magnetic system (SMS) project for the magnetic refrigerator machine is proposed. The second-generation high-temperature superconducting tapes for SMS are developed, fabricated and tested in NRC "Kurchatov Institute". The magnet consists of 12 non-insulated double pancake coils. The SMS is installed in the vacuum chamber and cooled by a cryo refrigerator.*

## 1. Introduction

Most of modern industrial refrigerators utilize the well-known gas-compression cycle. Nowadays this technology dominates, but it is not energy-effective enough. Moreover, increased environmentalists pressure to ban the use of harmful, toxic gases such as ammonia, CFC and others, makes it necessary to look for alternative cooling technologies [1]. The magnetic refrigerating machines (MRMs), in which the magneto-caloric effect is used, could be a good substitution for gas-compression refrigerators [1, 2]. By criteria of the profitability, MRMs have two branches of development. The first one deals with low-cost household systems with low energy consumption (below 1 kW) such as domestic refrigerators, air conditioning systems for vehicles, micro refrigerators for electronics and so on. For such devices, low cost and ease of exploitation are prioritized, so permanent magnets, such as NdFeB with magnetic field up to B = 1 T, are optimal for the magnetic system. The upper limit of the refrigerating capacity is 50 W per 100 $cm^3$ of volume.

The other branch of MRM concerns large-scale applications, such as industrial plants. The MRM refrigerating capacity grows almost linearly with increasing induction of the magnetic field *B* and the volume of the cooling magnetic material (regenerator), such as gadolinium. This causes interest in superconducting magnetic systems (SMS), which are capable to generate 3-5 T fields in the large volume. The disadvantage of the superconducting magnetic system is the need to maintain a zero resistance state, which imposes the requirement of cooling the winding to very low temperature. However, the use of high-temperature superconducting (HTS) tapes with the critical temperature ~ 90 K reduces the energy consumption down to 1-2 kW. Thus, the MRM with magnetic system fabricated from HTS tapes would be economically justified and promising for large-scale application with cooling power 10 kW and higher.

The usual MRM circle includes the following steps: the magnetization of a regenerator in the external magnetic field with transfer of generated heat to the thermostat; the demagnetization of regenerator accompanied by cooling down, what should be utilized by the heat exchange with refrigerated substance. The following main types of MRM based on the magnetization-demagnetization cycles are distinguished:

1. Reciprocating motion of the regenerator from the maximum to the minimum of the field;
2. Rotational movement of the regenerator between the poles of the magnet;
3. Periodic change of the external magnetic field with a fixed regenerator;
4. Rotation of the magnets between several regenerators.

For the large systems with a superconducting magnet, the last two types of MRM are not promising economically, since the energy costs of generating an alternating field are significant.

In the present work, the magnetic system design for MRM working near the room temperature is proposed. In the developed MRM the HTS tape fabricated in NRC "Kurchatov Institute" is used in SMS winding. In the paper the technical requirements for the superconducting tape are analyzed and the results of HTS tape current-carrying capacity investigation are given.

**2. HTS tapes fabrication**

The HTS tapes for magnet are manufactured at Superconducting Compact Pilot Line facility installed in Kurchatov Institute (Moscow). This facility is designed to study the technological principles employed in entire HTS processing chain. The tape length (~ 100 m) and width (4 mm) are sufficient to fabricate a double pancake coil for superconducting magnet. The architecture of the HTS tape is shown in Figure 1. HTS tapes with the same architecture have already demonstrated promising current performance in the high magnetic field [4] so this kind of tapes should be suitable for the magnet manufacturing.

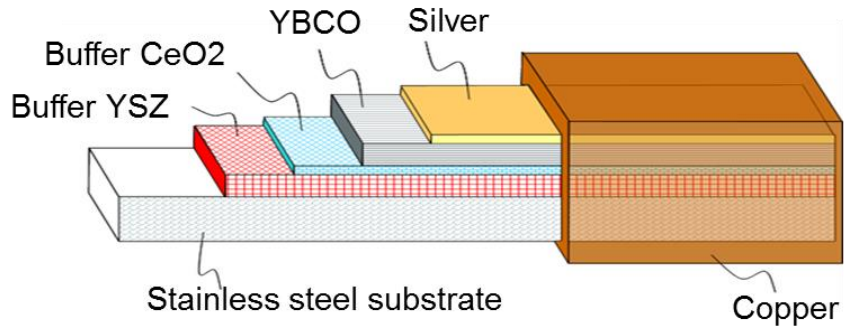

Figure 1. HTS tape architecture.

Two main deposition methods are used for the buffer and the superconductor layers fabrication. Alternating Beam Assisted Deposition (ABAD) method developed by Bruker [5] is used for fabrication ~2 μm thick yttrium-stabilized zirconia (YSZ) layer with a sufficiently high in-plane texture (with FWHM= 8.5–9.5°) on the stainless steel substrate (~ 100 μm). A high rate Pulsed Laser Deposition (HR-PLD) method developed in cooperation with Bruker [6] is used then for growing ~100-200 nm thick $CeO_2$ cap layer and ~1.5-2 μm thick $YBa_2Cu_3O_{7−δ}$ (YBCO) superconducting layer. The SEM image of tape cross-section demonstrates perfect quality of the functional layers (see Figure 2). The thickness of the silver coating layer is about 1-2 μm, the thickness of the copper plating is 10-20 μm on each side.

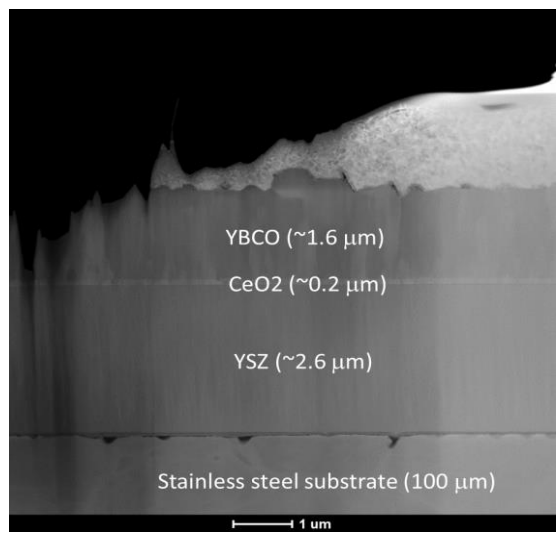

Figure 2. SEM image of HTS tape cross-section

These thicknesses can be adjusted to the required transverse conductivity of the pancake coils for the magnet. The total thickness of the tape is about 150 μm.

## 3. Investigations of the HTS tape properties

It is difficult to predict HTS tape critical current in high magnetic field and low temperature, based only on measurements at 77 K in the self-field. This is due to the complicated relationship between the value of the critical current and the specifics of the tape microstructure (including individual pinning structures). For example, the ratio of critical current density values at 4.2 K, 19 T, B||c, and at 77 K in the self-field, also known as "lift factor", may reach 10 in one kind of tapes, while other tapes give a lift factor of only 2–3 [5]. Therefore, the critical current of HTS tape should be estimated under field and temperature conditions close to the operating magnet conditions. For this purpose, both direct transport methods and magnetization measurements have been used.

The value of critical current measured by the transport method (77K, in self-field) is $I_c$ = 110 A. To test the uniformity of $I_c$ distribution over the tape area, reel-to-reel measurements by Tapestar have been used. The local $I_c$ is recalculated from the captured signals collected from the array of seven Hall sensors located perpendicular to the tape surface (at 77 K, external field of 8-10 mT). The uniformity of $I_c$ in the measured tapes is quite high. The deviation from the average is less than 10%.

The critical current decreases and demonstrates pronounced anisotropy in external magnetic field. Figure 3 shows the angular dependencies of the critical current in a magnetic field of 1.5 T at two temperatures (77 K – liquid nitrogen and 65 K – overcooled liquid nitrogen). The $I_c$ demonstrates a maximum at both temperatures when the field is directed along the tape plane (θ~90 deg., B||ab). A small $I_c$ maximum also appears when the field is

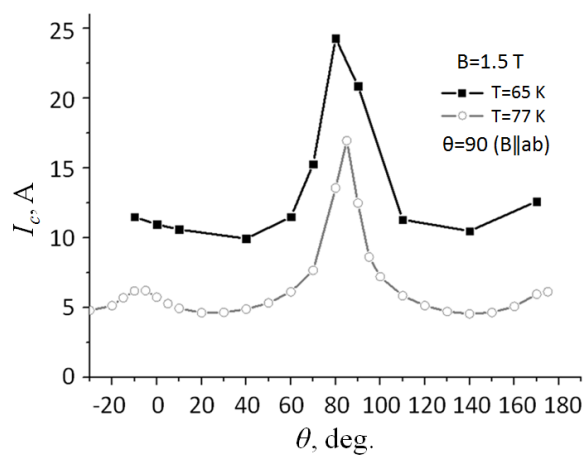

Figure 3. Angular dependencies of critical current in magnetic field of 1.5 T at 77 K and 65 K (DC measurements).

perpendicular to the tape plane (B||c). One can see that at liquid nitrogen and overcooled liquid nitrogen temperatures, the critical current is too low to use in magnet with 3 T - class field (engineering current density is only ~1700 A/cm$^2$ at 1.5 T).

At liquid helium temperature (4.2 K) the critical current of HTS tape increases considerably. In Figure 4, the measured transport critical current vs magnetic field dependence is shown in perpendicular orientation. At 4.2 K the critical current of HTS tapes in the field of 3 T exceeds 480 A, and it is about 300 A in the field of 10 T.

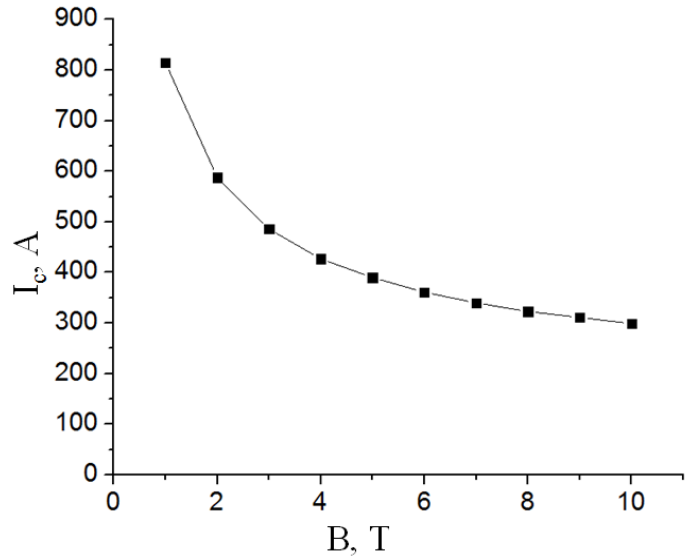

Figure 4.  Field dependence of HTS tape critical current in orientation perpendicular to the tape plane (*B*||*c*) at 4.2 K (transport measurements).

To estimate the acceptable operating temperature for SC magnet, magnetic measurements with Vibrating Sample Magnetometer Lake Shore VSM 7407 (up to 1 T) and Quantum Design XL7 MPMS SQUID magnetometer (up to 7 T) have been performed in range of temperature from 4.2 K to 77 K. The critical current of a small piece of tape (4~5 mm in size) is recalculated from the magnetization loops with the Bean model. Angular dependencies of relative critical current near perpendicular orientation in testing field 0.5 T at various temperatures are shown in Figure 5.

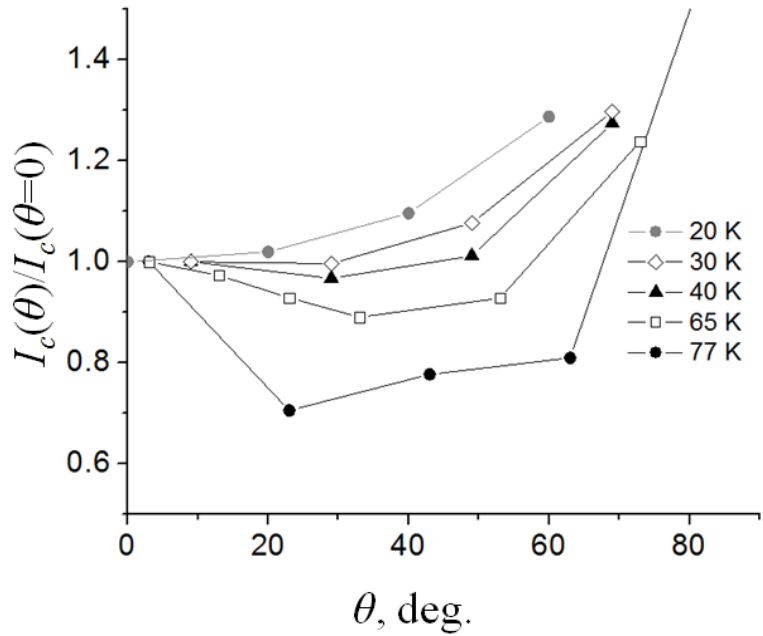

Figure 5. Angular dependencies of HTS tape relative critical current in magnetic field 0.5 T at various temperatures (magnetic measurements).

One can see that at temperature below 40 K in the field perpendicular to the tape plane (B||c) the critical current has a global minimum. No secondary $I_c$ maxima occur at other orientations except B||ab, which is consistent with other authors [7]. So B||c is the worst case of field orientation. The dependence of the critical current on the magnetic field in the B||c orientation is shown in Figure 6 for different temperatures. At this orientation, in magnetic field of 3 T the critical current exceeds 300 A at temperatures below 20 K.

Thus, taking into account superconducting properties of the HTS tapes and magnet design, the operating temperature should be in the range from 20 K to 30 K. Under these conditions, the value of safe transport current in 200 A is guaranteed for any orientation of the magnetic field.

**4. Magnet design**

The superconducting magnetic system (SMS) is designed in accordance with the requirements of the magnetic refrigeration machine. The magnetic field in the central part of the solenoid is about 3 T. The diameter of the warm (T ~ 300 K) bore is 60 mm. The solenoid winding height is $2b = 120$ mm. This value have been calculated from the requirement of the minimal displacement of magnetic regenerator with volume V = 400 cm$^3$ and taking into account the limitations of the one-piece of HTS tape length (60 - 100 m).

For the solenoid winding the double pancake structure is chosen. The main advantage of this structure is maintainability: in the case of quench the damaged pancakes could be easily replaced. In addition, this structure allows the use of pancakes with lower $I_c$, taking into account different orientations and values of the field in different coils.

The solenoid winding consists of 12 flat double pancake coils. Each of them is wound with the HTS tape manufactured at NRC "Kurchatov Institute". As mentioned above, this tape is a multilayer composite tape with 4 mm width and ~0.15 mm thickness. The double pancake is made of a single piece of HTS tape of length $l \sim 70$ m. Each double pancake has about 200 turns. There are cooling aluminum plates of ~ 2 mm thickness between the pancakes, so the effective thickness of the pancake is about 10 mm. The solenoid is characterized by the following parameters:

- $a_1 = 43$ mm, $a_2 = 60$ mm - internal and external radii of the coil;
- $2b = 120$ mm - height of the solenoid;
- $\alpha = a_2/a_1 \sim 1,4$; $\beta = b/a_1 \sim 1,4$;
- N = 200/cm – number of turns per 1 cm of height.

In this configuration, the ratio between the magnetic field $B$ and the current $I$ is defined as $B \sim 1.0 \cdot N \cdot I$ [8, 9], and the magnetic field in the central part is $B = 3.0$ T at $I = 150$ A.

The tapes of the solenoid windings are not insulated. It provides high radial thermal conductivity [10] and it is helpful for self-protection of the solenoid in the case of quench [11]. The critical current anisotropy must be taken into account as well. As it can be seen from Figure 3, the critical current has a maximum when the magnetic field is parallel to the tape, and it is considerably reduced when the angle between the normal to the tape and the field is less than $70^0$. At low temperature in the field 3-5 T, the anisotropy factor of $I_c$ in the longitudinal/transverse fields could reach ~ 4 [7].

It is well-known, that the magnetic field lines at the center of the solenoid are parallel to its axis, and curved at the edge. Figure 7 shows the magnetic field lines at the edge of double pancake, located at exterior of the solenoid. It can be seen that the maximum field B = 2.5 T at the edge of the solenoid is directed at an angle $\theta = 55\text{-}60^0$ to the normal of the tape. According to the anisotropy $I_c(B,\theta)$ shown in Fig. 3 and Fig. 5 as well as in ref. [7], the critical current is considerably reduced compared with the situation when the magnetic field is parallel to the tape surface. To increase the operating current of the solenoid (and, therefore, the magnetic field), the operating temperature must be reduced. In Figure 6, the dashed line shows the characteristic function of the solenoid B(I). In the temperature range of 20–25 K, the critical current is substantially higher than the operating current at I = 150 A for a 3 T field.

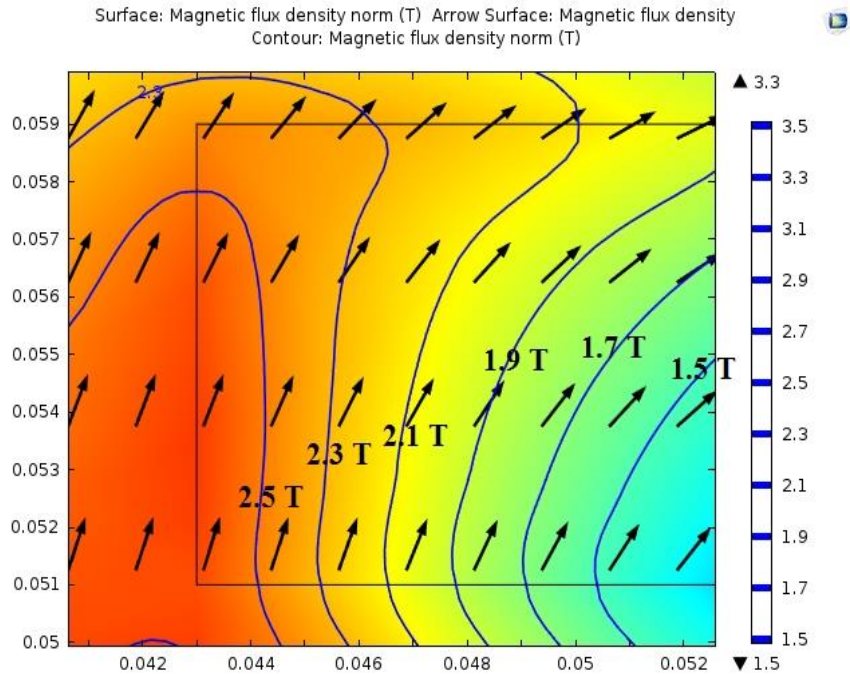

Figure 7. The map of the magnetic field at the edge of the exterior double pancake. The solenoid axis is on the left. The arrows are showing the direction of the magnetic field in corresponding points. Solid curves represent lines with constant modulus of the magnetic field.

The pancakes are cooled by aluminum (AD-1) plates with high thermal conductivity [12]. The plates have 2 mm in thickness and placed between pancakes. The surfaces of these plates are anodized. The thickness of the oxide layer $Al_2O_3$ on the plate's surface is about 15 μm. Thin dielectric layer of oxide provides electrical insulation of the pancakes. The thermal contact of the plates and the pancakes is ensured by Apeizon N vacuum grease, which doesn't affect the critical parameters of the tape and eliminates the problem of delamination [13].

The left side of Fig. 8 shows the superconducting magnetic system, cooled by the cryocooler (3). The solenoid itself (4) is shown on the right side of Fig. 8. The solenoid (4) is installed on the copper plate of the thermal shield (2) using a "conical saw" made of textolite (fabric-based laminate), which provides thermal insulation. The copper plate is installed on five supports (5) of textolite tubes with a diameter of 8 mm, a height of 50 mm and a wall thickness of 2 mm. Three stainless sweep steel strings (6) rigidly fix the solenoid to the bottom flange, which provides mechanical stability. At a height of ~70 mm, the strings are soldered to the thermal shield.

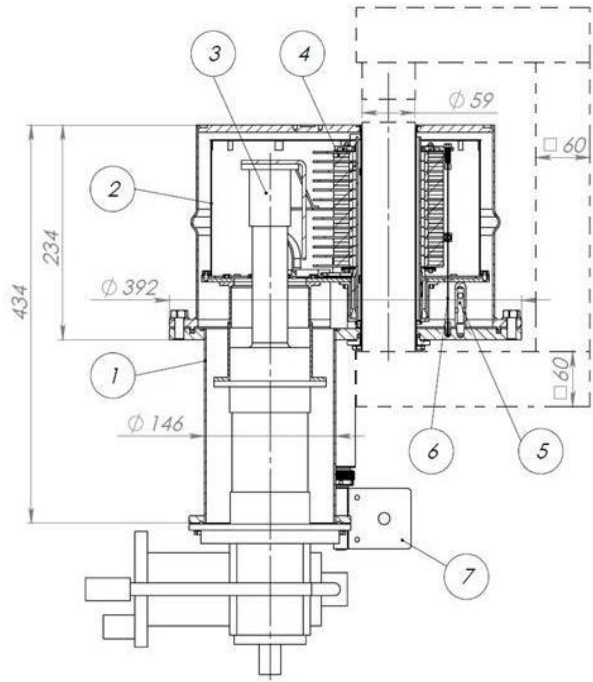

Figure 8. Left: superconductor magnetic system with cryocooler. 1- vacuum jacket, 2 - thermal shield of the 1-st stage, 3 - cryocooler, 4 - superconducting solenoid, 5 - shield support, 6 - fixing string, 7 - current lead. Right: solenoid construction from double pancake windings.

The solenoid placed as close as possible to the top of the vacuum jacket (1). This is necessary because it is planned to use an external yoke, made from the soft ferromagnetic material (permendur), which is shown by a dotted line in Fig. 8. The estimations have shown that, although the magnetic field decreases outside the solenoid, the permendur yoke allows generating the field of 3 T in an interpolar air gap of 20-25 mm. This space is planned to be used in rotational option of magnetic refrigeration machine.

## 5. Heat calculations

The operating temperature of the solenoid cooled by the 2-nd stage of the cryocooler is $T_2 = 20\text{-}25$ K. The temperature of the 1-st stage is $T_1 = 40\text{-}45$ K. The used model of the cryocooler RKD-408D2 provides cooling capacity of 35-45 W in the 1-st stage and 15 W in the 2-nd stage, respectively. The proposed concept of SMS is designed so that the heat flows to corresponding stages do not exceed the cooling capacity of the cryocooler.

The heat input to the cryocooler is due to the supports, the measuring wires and the current leads. The thermal shield is cooled by the 1-st stage and is heated by the heat radiation from the vacuum skirt of the cryostat. Thermal radiation to the second stage could be neglected.

The heat leak to the supports is estimated by the well-known ratio for a massive conductor with a constant cross section:

$$Q = \frac{S}{l} \cdot \int_{T_0}^{T_1} \lambda(T) dT \ [W] \tag{1}$$

where $S$ - cross section of the conductor [m$^2$], $l$ - the length of the conductor [m], $T$ – temperature [K]. The coefficient of thermal conductivity $\lambda(T)$ [W/m·K] was determined by interpolation of the tabular data for the heat-conducting material (stainless steel, copper, brass, textolite etc.) [14]. The calculated heat input on five supports does not exceed $10^{-3}$ W.

Due to the large temperature difference between the vacuum jacket ($T_0 = 300$ K) and the copper shield ($T_1 = 45$ K), a noticeable heat flux is generated. The influx of heat due to radiation was estimated from the well-known expression:

$$Q_{01} = \varepsilon_{01} \cdot \sigma \cdot (T_0^4 - T_1^4) \cdot S \ [W] \tag{2}$$

$\sigma$ is the Stefan-Boltzmann constant, $T_0$, $T_1$ are the shell (jacket) and the body (shield) temperatures. The reduced radiation coefficient $\varepsilon_{01}$ is [14]:

$$\varepsilon_{01} = \frac{1}{\frac{1}{\varepsilon_0} + (\frac{1}{\varepsilon_1} - 1) \cdot \frac{S_0}{S_1}} \tag{3}$$

where $\varepsilon_1$, $\varepsilon_0$ are the radiation coefficients of the material of the body and the shell, $S_1$, $S_0$ - surface area of the body and the shell.

For stainless steel $\varepsilon_0 = 0.075$ at 300 K and $\varepsilon_1 = 0.07$ for copper at 40 K. According to (2), the heat transfer to the flat surfaces of the cryostat thermal shield is 2.64 W, and to the lateral surfaces the heat inflow is 4.37 W.

### 6. Current leads

Each current lead of the solenoid consists of two sections. The first section is made of copper and extended from room temperature to the 1-st stage of the cryocooler (~ 45 K). The second section is made of HTS tapes - from the 1st stage of the cryocooler to the solenoid terminals. Calculations and optimization of uncooled current leads are discussed in detail in [15, 16]. For copper, the optimal ratio between the current *I*, the length *L* and the cross section *S* is:

$$\frac{LI}{S} \approx 4 \cdot 10^4 \; [\frac{A}{cm}] \tag{4}$$

In accordance with this ratio for $L = 25$ cm and $I = 200$ A the cross section should be $S \sim 12.5$ mm$^2$. In the absence of current, the heat input to the first stage is $Q \approx 4.9$ W. Taking into account the temperature distribution over the current lead [16] at an operating current of 150 A, $Q \approx 7$ W per one current lead. Thus, the total heat inflow to the 1-st stage of the cryocooler with the operating current 150 A is $Q \approx 21$ W.

To calculate the heat input to the 2-nd stage, the following assumptions are made: the temperature of the 2-nd stage is $T_2=20$ K, the temperature of the 1-st stage is $T_1 = 45$ K. Due to the small temperature difference, the heat transfer due to radiation is small, it does not exceed 25mW.

The main part of heat generation occurs when the solenoid is operating. The typical resistance of soldered contacts between the tapes of pancakes and superconducting current lead does not exceed 0.1µOhm. The total number of such contacts is 26 and, therefore, at current $I = 200$ A, their heat generation does not exceed 100 mW. The heating of the winding during the current change depends on the rate and will be determined experimentally.

The heat transfer along the strings to the top of the solenoid does not exceed 0.05 W. The heat input due to 20 measuring leads (manganin, diameter 0.05 mm) is about 5 mW. The total heat transfer to the 2-nd stage of cryocooler over two HTS current leads is 200 mW [16]. Thus, the total heat transfer to the 2-nd stage of the cryocooler is less than 400 mW.

The estimations of heat flows show that the refrigerating capacity of the 2-nd stage of this cryocooler at T= 20-25 K is sufficient to provide the stable operation of the solenoid.

### 7. Conclusions

This paper proposes an innovative concept for the superconducting magnetic system of a magnetic refrigerating machine. Corresponding technical specifications for HTS tapes are formulated. The current-carrying capacity of HTS tapes fabricated in NRC "Kurchatov Institute" is investigated. It is shown that HTS tapes can be used to manufacture solenoids with a magnetic field of 3 T and an operating temperature of 20-25 K.

**Acknowledgments**

The authors thank Dr. A.V. Emelyanov, Dr. N.K. Chumakov for discussions of the results, NBICS-NlT Resource Center of Electro-physical Methods for the HTS tapes magnetic measurements and NBICS-NlT Resource Center of Probe and Electron Microscopy for SEM investigation.

This work was supported by the Ministry of Science and Higher Education of the Russian Federation with agreement No.14.604.21.0197 (the unique identifier of applied scientific research RFMEFI60417X0197).